\documentstyle[12pt]{article}

\begin{document}

\title{\bf{A Stochastic Approach to the Construction of One-Dimensional Chaotic
Maps with Prescribed Statistical Properties}}

\author{F.K. Diakonos\\
Department of Physics, University of Athens\\ \vspace*{0.5cm}
GR-15771, Athens, Greece\\ 
D.Pingel and P.Schmelcher\\
Theoretische Chemie, Physikalisch--Chemisches Institut\\
Universit\"at Heidelberg, INF 229, D-69120 Heidelberg\\
Federal Republic of Germany}

\date{\today}
\maketitle








\begin{abstract}
We use a recently found parametrization of the solutions of the inverse
Frobenius-Perron problem within the class of complete unimodal maps 
to develop a Monte-Carlo approach for the construction of one-dimensional
chaotic dynamical laws with given statistical properties, i.e. invariant
density and autocorrelation function. A variety of different examples are presented
to demonstrate the power of our method.
\end{abstract}


\section{Introduction}

Chaotic dynamical systems show a rich diversity of possible behaviour with respect
to their statistical properties. In recent years much work has been devoted to 
the understanding of how these statistical properties emerge from the
dynamics. The most popular area for such investigations are
unimodal 1-d maps \cite{Col80,DynSta,Gros87,Cso93}. They became widespread due to two 
main advantages: their dynamics can be calculated efficiently enough to make
extensive numerical investigations and the variety of statistical properties
within this class of dynamical laws is very large.
Certain features of special systems can even be investigated analytically.
The interplay between the dynamical and statistical behaviour becomes more
transparent if we consider the inverse problem, i.e. given the 
statistical behaviour of a system how could we extract relevant information on the 
possible dynamics. This subject has recently been addressed and discussed in the literature
\cite{InvPro,Dia96,PiSchmeDia,BaraDa,Bollt,Gora} and from a communications point of view
in refs. \cite{Abel,Broom}.
The purpose is then to construct an one-dimensional fully chaotic (and ergodic) dynamical
law for a given invariant density or/and time autocorrelation function.
Given the latter we can on the one hand calculate all expectation values,
i.e. statistical averages, of measurable observables depending exclusively
on the single dynamical variable. On the other hand the given autocorrelation function
provides us with valuable information on the dynamics of the
trajectories of the system. It represents an important quantity strongly related to the
physics described by the dynamical system.
The construction of a dynamical system with prescribed statistical
properties can have the background of either some experimentally given data for which a dynamical
system should be modelled or it can be justified by the fact that
a definite behaviour of the correlation function is required in the
context of the control of a physical system \cite{Bollt,Gora}. 

Restricting ourselves to the first part of the problem,
i.e. the construction of a map with a given invariant
density (the so called inverse Frobenius-Perron problem (IFPP)) we have 
recently found a general and for practical purposes very helpful representation
of the solution of the IFPP within the class of smooth complete and unimodal maps
\cite{PiSchmeDia}. However the combined problem (both the invariant density and time autocorrelation
function are given) is much more complicated. Some progress in this direction
can be made if one uses the class of piecewise linear Markov maps as a 
basis for this search. Constructing such a map with a given
autocorrelation function one can then obtain the proper invariant density
performing a suitable conjugation (coordinate) transformation \cite{BaraDa}. A substantial
disadvantage of this method is that the changes of the autocorrelation function
in the course of the conjugation transformation are not fully under control.
Another rather anaesthetic aspect is the fact that in many cases the
solution found to the inverse problem does not fulfill certain
smoothness criteria.

The purpose of this letter is to present a new approach to the inverse
problem including the autocorrelation function.
It is based on the general representation of solutions of the
IFPP found in \cite{PiSchmeDia} for complete smooth unimodal maps.
We suitably parametrize the key function $h_f$ occurring
in this approach and develop a stochastic
(Monte-Carlo) algorithm to determine the optimal solution such that the
$\chi^2$ deviation of the resulting autocorrelation from a 
desired function is minimized. The paper is organized as 
follows: In section 2 we briefly review the representation of our solution
to the IFPP. In section 3 we present the Monte-Carlo algorithm used to perform the 
$\chi^2$ minimization with respect to the time autocorrelation function. 
Finally in section 4 we provide several applications in the framework of our Monte-Carlo 
scheme and discuss the convergence properties of our approach.

\section{The general solution of the IFPP for smooth unimodal 1-d maps}

In a recent paper \cite{PiSchmeDia} we investigated the problem of designing
dynamical systems which possess an arbitrary but fixed invariant density.
The solution to this inverse problem is of great interest for the numerical
simulation of real physical systems as well as for the understanding of the
relationship between the functional form of the map and the statistical
features of the corresponding dynamics. We were able to derive a general
representation of all ergodic and chaotic complete unimodal maps with a given
invariant density. This corresponds to the general solution of the so called
inverse Frobenius-Perron problem
\cite{DynSta,InvPro,BaraDa,Gross,Ghik,Pala,Hunt}
within this class of maps. We review here the basic aspects of our
approach.

As a starting-equation for the construction of the map we use the
Frobenius-Perron equation:
\begin{equation}
\rho(y)\,|dy|=\sum\limits_{x_i=f^{-1}(y)}\rho(x_i)|dx_i|
\label{froperequ}
\end{equation}
where the summation runs over all preimages of $y$ (for unimodal maps
$i=L(left),R(right)$).
For any given complete unimodal $f(x)$ the right preimage of $y$ is
determined if the left preimage is given and vice versa.
The essential feature of our approach is the following: A
prescribed relation between the two preimages reduces the number of
independent differentials on the rhs of eq.(\ref{froperequ}) and allows 
to integrate the Frobenius-Perron equation.
Such a relation is given by the function $h_f(x)$ which maps the position of
the left preimage onto the position of the right one:

\begin{eqnarray}
h_f&:&[0,x_{\max}]\longrightarrow[x_{\max},1] \nonumber \\
x_R&=&h_f(x_L)\;\;\;\;\;\;\mbox{with}\;\;f(x_L)=f(x_R)
\label{hfprei}
\end{eqnarray}
where $x_{\max}$ is the position of the maximum of the map.
$h_f(x)$ is a monotonously decreasing function on the defining interval
and it is differentiable with the exception of a finite number of 
points. It obeys the equations:

\begin{eqnarray}
h^\prime_f(x)<0 &&\;\;\;\;x\in[0,x_{\max}] \nonumber \\
h_f(0)=1  &&\;\;\;\;h_f(x_{\max})=x_{\max}
\label{hfprop}
\end{eqnarray}
In terms of $h_f$ we obtain the general solution to IFPP within the class
of unimodal maps as:
\begin{equation}
f(x)=\mu^{-1}\left(1-|\mu(x)-\mu(H_f(x))|\right)
\label{funx}
\end{equation}
where $H_f(x)$ is:
\begin{equation}
H_f(x)=\left\{ \begin{array}{l@{\quad;\quad}l}
        h_f(x)&0 \leq x<x_{\max}\\
        h^{-1}_f(x)&x_{\max}\leq x \leq 1\\
        \end{array}\right.
\label{genhf}
\end{equation}
and $\mu(x)=\displaystyle{\int_0^x} \rho(y) dy$ is the corresponding
invariant measure.
All unimodal maps with prescribed invariant density $\rho(x)$ are given by
(\ref{funx}), $h_f(x)$ taking on all possible functional forms obeying
(\ref{hfprei}) and (\ref{hfprop}) (for more details on the above derivation
as well as its relation to the standard conjugation procedure we refer
the reader to \cite{PiSchmeDia}).

The above shows that fixing the invariant
measure is a relatively weak constraint in the framework of the inverse problem and
that there is still a considerable freedom to model the mapping.
In the next section we will use this freedom and suitably parametrize
$h_f(x)$ such that, by tuning appropriately the corresponding parameters,
we get a map with an autocorrelation function possessing a minimum deviation
(to be defined below) from a desired correlation function.

\section{Stochastic optimization as a tool for the construction of a map}

Before presenting our search method for the optimal map in the sense of
some desired statistical properties let us specify in more detail 
what we want to achieve and what our input for the problem at hand is.
Consider a given invariant density $\rho(x)$ (corresponding to a finite
measure) which arises in the asymptotic limit ($t \to \infty$) from the
dynamics of some unknown $1-d$ fully chaotic single humped map. Consider also
as given the first $m$ values $\{C(1), C(2),..,C(m)\}$ of the time
autocorrelation function $C(n)$
(note that $C(0)$ is determined entirely through $\rho(x)$).
We are seeking a map  $f(x)$ which possesses the above statistical
properties ($\rho(x), C(n) (n=0,...,m)$). We focus in the following on the
representation (\ref{funx}) for all admissible maps where the auxiliary
functions $h_f(x)$ fulfill the requirements (\ref{hfprei},\ref{hfprop}).
Using this expression for $f(x)$ it is guaranteed
that the map we are looking for possesses
the invariant density $\rho(x)$ (provided of course that  
$\mu(x)=\int_0^x \rho(t)~dt$ in (\ref{funx}) exists).
All our freedom in modelling the map $f(x)$ is now contained in the auxiliary 
function $h_f(x)$. We will use a suitable representation of $h_f(x)$
to parametrize the map $f(x)$. This parametrization is the starting
point for a stochastic optimization procedure to obtain
a map with the desired autocorrelation function $C(n)$ at times $n=1,2,..,m$.
One of the simplest ansatz would be to write down a piecewise linear 
expression for $h_f$. Our knowledge of the properties and dynamics of $1-d$ maps
however suggests that the local behaviour of the map in the neighbourhood of
special points, like for example a marginal unstable fixed point in the case
of intermittent dynamics (see ref.\cite{Ott} and references therein), plays a crucial role in
determining the statistical properties of the map. A more intuitive
ansatz is therefore needed. Here we will proceed as follows.\\
We define a $1-d$ lattice with $N+1$ points in the interval $[0,x_{\max}]$.
The coordinates of the lattice points are $x_p(i),~i=0,1,..,N$ with
$x_p(0)=0$ and $x_p(N)=x_{\max}$. The function $h_f(x)$ is piecewice defined.
We will call in the following the expression of $h_f$ in the $i-th$ interval
$[x_p(i-1),x_p(i))$ an {\bf{element}} and use the notation
$h_{f,i}$ for the $i-th$ element. In practice one is free to choose many
different expressions for the $i$-th element. Constraints on $h_f$
as for example continuity in $[0,x_{\max}]$ together with the conditions
(\ref{hfprei},\ref{hfprop}) however restrict the possible forms.
A desirable but not necessary requirement is that $h_{f,i}$ should be analytically
invertible in order to easily extend the solution in $[0,x_{\max}]$ to
the interval $(x_{\max},1]$ (see (\ref{genhf})).
Here we will investigate two different choices for the element $h_{f,i}$.
In both cases continuity of $h_f$ is guaranteed.
The first case, refered to in the following as model I, is obtained by choosing the 
order of $h_{f,i}$ at the left point $x_p(i-1)$ in the interval
$[x_p(i-1),x_p(i))$. This parametrization allows us to fit the derivative
of the map at $x=0$ and has the advantage of using a minimal set of
parameters in describing the element $h_{f,i}$. The disadvantage of this
choice is that one cannot fit independently the order of the maximum of the
map at the right point $x_p(N)$ of the last interval $[x_p(N-1),x_p(N)]$.
This is the reason for considering also a second case (model II). Here
we use for the element $h_{f,i}$ an expression representing expansions around
both limiting points $x_p(i-1),x_p(i)$ which match together
at some point of the interval $[x_p(i-1),x_p(i)]$. An additional
parameter is therefore required in order to satisfy continuity for $h_{f,i}$ and its
first derivative in $(x_p(i-1),x_p(i))$. Due to this fact model II uses
2 more parameters than model I for the description of a single element $h_{f,i}$.

Let us now discuss the two models in more detail. For model I the element
$h_{f,i}$ is given by:
\begin{eqnarray}
h_{f,i}&=&(y_p(i)-y_p(i-1)) \left(\frac{x-x_p(i-1)}
{x_p(i)-x_p(i-1)}\right)^{\alpha(i)} + y_p(i-1)
\nonumber \\
~~~x&\in&[x_p(i-1),x_p(i))~~;~~~i=1,2,..,N
\label{hfexp}
\end{eqnarray}
where $y_p(i-1),y_p(i)$ are the values of $h_f$
at the points $x_p(i-1),x_p(i)$ respectively,
while the power $\alpha(i)$ determines the local behaviour of $h_f$
around $x_p(i-1)$. Due to the constraints (\ref{hfprei},\ref{hfprop}) we get:
$y_p(0)=1$ and $y_p(N)=x_{\max}$. The monotony of $h_f$ implies that 
$y_p(i-1) > y_p(i)$ for $i=1,2,..,N$. The above ansatz determines the map
$f(x)$ in the interval $[0,x_{\max}]$. To find $f(x)$ in the remaining
interval $[x_{\max},1]$ we take advantage of eq.(\ref{funx}) and therefore
have
to invert the auxiliary function $h_f(x)$. It is straightforward to show that the expression
(\ref{hfexp}) can be inverted analytically leading to a closed form for
$h_f^{-1}(x)$. The above defined ansatz for $h_{f}$ is in general piecewise
smooth, i.e.
smooth with the exception of a finite number of points.

In model II we use the following ansatz for the element $h_{f,i}$:
\begin{eqnarray}
h_{f,i}(x)&=&y_p(i-1) - c_L(i) (x-x_p(i-1))^{\alpha_L(i)}~;~
x \in [x_p(i-1),x_s(i)) \label{hf2eq1}\\
h_{f,i}(x)&=&y_p(i) + c_R(i) (x_p(i)-x)^{\alpha_R(i)}~~~~;~~~~
x \in [x_s(i),x_p(i)] \label{hf2eq2}
\end{eqnarray}
where we have introduced four new parameters
$c_L(i),c_R(i),\alpha_R(i),x_s(i)$
and a new point $x_s(i) \in [x_p(i-1),x_p(i))$ for each $i=1,...,N$.
Two of these parameters, namely $c_L(i)$ and
$c_R(i)$, can be fixed demanding continuity of $h_f$ and its first derivative
at $x=x_s(i)$. We arrive then at the following generating formula for $h_f$:
\begin{eqnarray}
h_{f,i}(x)=y_p(i-1) - &S_i& \alpha_R(i)(x_s(i)-x_p(i-1))^{1 - \alpha_L(i)}
 (x-x_p(i-1))^{\alpha_L(i)}~; \nonumber \\
 x &\in& [x_p(i-1),x_s(i)) \nonumber \\
h_{f,i}(x)=y_p(i) +
&S_i& \alpha_L(i)(x_p(i)-x_s(i))^{1 - \alpha_R(i)}
(x_p(i)-x)^{\alpha_R(i)}~; \nonumber \\
x &\in& [x_s(i),x_p(i))
\label{hfexpn}
\end{eqnarray}
with:
\begin{equation}
S_i=\frac{(y_p(i-1)-y_p(i))}
{\alpha_R(i) (x_s(i)-x_p(i-1)) + \alpha_L(i) (x_p(i)-x_s(i))}
\label{shel}
\end{equation}
$h_f(x)$ given in eq.(\ref{hfexpn}) can also be inverted
analytically to obtain $h_f^{-1}$ defined in $[x_{\max},1]$.
Comparing eqs.(\ref{hfexp},\ref{hfexpn}) we see immediately that the element
of model II has 2 more parameters than the element of model I.

In the remaining part of this section we will discuss the algorithm
which allows us to obtain dynamical systems possessing certain statistical
properties based on the above ansatz for the auxiliary function $h_f(x)$.
To this end we will concentrate on the parametrization of model I.
One can then directly apply these ideas to the case of model II. 
We have determined the ansatz of $h_f(x)$ and therefore also of $f(x)$
in terms
of the parameters $\{x_p(i),y_p(i),\alpha(i)\}$. Once the values of these
parameters are chosen the map is completely specified. Due to its appearance
(see eq.(\ref{funx})) we automatically know its invariant density (measure)
and expectation values of observables. The corresponding
autocorrelation function
can be obtained via its defining formula
\begin{equation}
C(n)=\int_{0}^{1}x f^{(n)}(x) d\mu(x) - \left(\int_{0}^{1}x d\mu(x)\right)^2
\label{corfct}
\end{equation}
which requires a numerical integration. Here $f^{(n)}$ is the $n-th$ iterate
of the map $f$.

To proceed with our central subject, the inverse problem, we assume
that the first $m$ values of the autocorrelation function are given.
This can be due to some experimentally given data for which a dynamical
system should be modelled or due to the fact that
a definite behaviour of the correlation function is required in the
context of the control of a physical system. As discussed above each set 
$\{x_p(i),y_p(i),\alpha(i)\},~i=1,2,..,N$ respecting the constraint
of monotony determines uniquely the 
autocorrelation function $C_{h_f}(n)$ of the corresponding map $f(x)$.
One can now ask for the best set $\{x_p(i),y_p(i),\alpha(i),i=1,...,N\}$ in
the
sense that the resulting $C_{h_f}(n)$ possesses the least possible deviation
from the given autocorrelation $C(n)$.
As a measure for the above-mentioned deviation one can use a $\chi^2$-like
cost function. We therefore introduce the following quantity:
\begin{equation}
K[h_f]=\sqrt{\sum_{j=1}^{m} \left( \frac{C_{h_f}(j)-C(j)}{C(j)} \right)^2}
\label{chi2}
\end{equation}
The functional $K[h_f]$ is a highly nonlinear function of the
parameters $\{x_p(i),y_p(i),\alpha(i)\}$ and we are looking for the global
minimum of this function. To perform the minimization of $K[h_f]$ we
use a Monte-Carlo (MC) approach based on the Metropolis algorithm
\cite{Metro}.

The minimization is performed in several steps, increasing in each step
the number of elements $h_{f,i}$ used for the determination of $h_f$.
We start with a lattice consisting of only two points 
(the origin $x_p(0)=0$ and the position of the maximum $x_p(1)=x_{\max}$).
This means that only one element $h_{f,1}$ is needed for the specification
of $h_f$. The parameters to be fitted in this case are only two:
$x_{\max}$ ($x_p(1)$) and the power $\alpha(1)$ determining the behaviour of
$h_f(x)$ in the neighbourhood of the origin. 
The first step ends when the MC minimization has converged to some optimal
values for the two fit-parameters.
In the second step we use a lattice with three points $x_p(0),x_p(1),x_p(2)$
with $x_p(0)=0,~x_p(1)=\frac{x_{\max}}{2},~x_p(2)=x_{\max}$. Now we need
two elements $h_{f,1}$ and $h_{f,2}$ to determine $h_f$.
We do not keep the values of the old fit-parameters ($x_{\max}$,$\alpha(1)$)
fixed in the second step. Instead we use 
Gaussian distributed random variables for the choice of the old
fit-parameters. The mean values of these Gaussians are 
the optimum values obtained for these parameters in the previous step and
the corresponding widths are taken small enough to allow only weak
fluctuations ($\approx 10\%$) around the mean values.
Again we perform a MC optimization to obtain optimal values for
the two new parameters. This procedure is repeated until the desired convergence is achieved.
In each step the lattice size is increased by one point while we include two new parameters.

We use the Metropolis algorithm to find
the optimal values for $x_p(i)$ and $\alpha(i+1)$ in each step.
For every trial in the $i$-th turn we assume that
$x_p(i)$ and $\alpha(i+1)$ follow a uniform distribution in $(0,1)$ and
$(0,\infty)$, respectively (in practice the interval $(0,\infty)$ is replaced
by the finite one $(0,c)$ with an upper cutoff $c>>1$).
The annealing is introduced
through a thermalized probability distribution of the type: $P=e^{-K[h_f]/T}$
to avoid the trapping into local minima. The parameter $T$, playing
the role of the temperature, is positive and has to be tuned
adiabatically to smaller and smaller values such that the global minimum is
reached asymptotically.

We do not keep the values of the old fit-parameters fixed in the following step.
Instead we relax this constraint using
Gaussian distributed random variables for the choice of the old
fit-parameters. The mean values of these Gaussians are 
the optimum values obtained for these parameters in the previous step and
the corresponding widths are taken small enough to allow only weak
fluctuations ($\approx 10\%$) around the mean values.

Adding new fit parameters allows us to determine $h_f$ in more and more detail and improves
the convergence of the autocorrelation function of the model dynamical system to the given
(experimental) autocorrelation function. Indeed, as we shall see below, only a few
elements $h_{f,i}$ are required to achieve a rather good convergence.
In the next section we give some
examples demonstrating how the above-described method can be applied to
construct a (piecewise) smooth map simulating a system with given time correlations. 

\section{Numerical examples and discussion}

Before turning to the examples and results of our computational method let us
provide some additional aspects concerning the determination of the
correlation function.
We restrict our investigations to
a rather small set of values for the correlation function $C(n)$,
more precisely
to the set $\{C(0),...,C(5)\}$. The reason for this restriction is that the exact
calculation of the correlation function which has to be accomplished for each single step
of the Monte-Carlo scheme is computationally very intensive (see below,
the total amount of CPU of our calculations on a powerful workstation was approximately
three months). The reliable evaluation of the correlation function is
by no means trivial.  As demonstrated for example in ref.\cite{SPDan}
the results obtained for the correlation function calculated with the
trajectories of the dynamical system are, in many cases, not reliable and 
cannot be improved by going to longer propagation times. Therefore
other methods for the calculation of the time
correlations are needed. Here we use a numerical approach to eq.(\ref{corfct}).
It is based on the extraction of the monotony intervals for the $n$-th iterate.
The endpoints of the monotony intervals are given as the preimages of the
maximum $x_{\max}$. The integration is then performed for each monotony 
interval separately. This ensures an accurate although very CPU time consuming
evaluation of the correlation function. The latter is related 
to the exponential growth ($2^n$) of the number of monotony intervals with
increasing $n$.

We apply now the stochastic method described in the previous section to several examples
to demonstrate its capability of providing dynamical systems with
prescribed statistical properties. The four different cases we study here
can be summarized with respect to their statistical properties as follows:
\begin{itemize}
\item{exponentially decaying autocorrelation function and uniform 
density}
\item{exponentially decaying autocorrelation function and linear 
density}
\item{oscillatory decaying autocorrelation function and power-law density} 
\item{power-law decaying autocorrelation function
and  uniform density}
\end{itemize}

Figs.(1-4) illustrate the corresponding results of our Monte-Carlo approach.
The subfigures (a,c) within each figure show the resulting maps for model I and
model II, respectively. Each of the subfigures (b,d) within each figure (1-4) shows
both the prescribed data for the autocorrelation signal as well as the result of our
optimization approach for model I and II, respectively.
In comparing the given autocorrelation data and the results of
our optimization scheme we deduce in the following a
'mean relative error' per individual data point $\frac{K[h_f]}{\sqrt{m}}$
of the autocorrelation function.

Since it is not our goal to provide as precise data as possible for the
optimized autocorrelation functions but to demonstrate the feasibility
of our approach with a good accuracy for a few data points we restrict
ourselves to auxiliary functions $h_f$ composed of a few elements $h_{f,i}$.
For the case of model I we use (with one exception) four points, i.e.
three elements $h_{f,i}$ for the
decomposition of the auxiliary function $h_f$. For model II we use
only two points, i.e. one element $h_{f,i}$.
Typically a few ten thousand Monte-Carlo steps are performed.
The most time consuming part of the algorithm is the calculation of the
autocorrelation function which has to be done for each MC step.

First we consider a simple example for which the maps we are looking for are
well-known. We use an exponential decay as a prescribed behaviour for the 
autocorrelation signal while the invariant density of the 
dynamical process is assumed to be uniform in $[0,1]$. A corresponding family of maps,
the nonsymmetric tent maps, 
has been studied in some detail in ref.\cite{Mori}. They are the nonsymmetric tent maps.
The resulting maps of our Monte-Carlo optimization are illustrated for model I in fig.1(a) and
for model II in fig.1(c). They show only minor differences, i.e. the outcome
of approach I is almost the same as for approach II, and both
are to a very good approximation an asymmetric tent map.
Both models lead also to a very good approximation of the corresponding prescribed correlation
function with an error of only $2.2\%$.

For the second example the two models lead to quantitatively different
results. The invariant density is supposed to be linear while the
autocorrelation function shows an exponential decay similar to the previous case.
The results of the stochastic minimization for this case are illustrated in figure 2.
The two obtained dynamical systems (figs.2(a,c)) have a very different
appearance reflecting the fact that our modelling procedure allows for
various dynamical systems with the same invariant density and autocorrelation.
Let us briefly address the main features of the resulting maps.
The map of model I possesses an almost marginal unstable fixed point at the origin
and an almost vertical derivative at $x=1$. Additionally it possesses
two obvious points of noncontinuous derivatives located on the left and right branch
of the map, respectively. The one on the left branch has a nonvanishing right derivative
which, via the function $h_f(x)$, is mapped to a point with almost vertical left derivative
in the right branch of $f(x)$. The reader should note that the latter point
coincides with the nonzero fixed point of the map !
The map of model II (fig.2(c)) is almost straighlined on its left branch and shows
an almost vertical right derivative at the maximum and an almost vertical left
derivative at $x=1$. In general the map of model II is 'smoother' than
the map of model I, which is an overall tendency to be observed
in any of our examples. It can be viewed as a result of the additional flexibility
within model II which allows to independently adapt the left and right derivative
in a given interval thereby joining them smoothly together (see the above description
for the ansatz of $h_f$ in model II).
The prescribed and optimized autocorrelation data for model I are illustrated in subfigure 2(b): 
They show a deviation of $129 \%$ which is predominantly due to the inability of reproducing
the single point $C(1)$ within the approach of model I.
Although hardly visible in fig.2(b) the prescribed and optimized data
coincide very well for $\{C(n),2\le n \le5\}$.  The optimized correlation function for
model II leads to a much better approximation to the prescribed
exponential decay and yields an error of only $5.4 \%$ (see fig.2(d)).

As a third example we use an invariant density obeying a power law with
an exponent $\beta=-0.5$ while the autocorrelation function is chosen to
possess a decaying oscillatory behaviour.
The corresponding minimization results are shown in fig.3.
Both maps (figs.3(a,c)) show an almost horizontal derivative
at $x=1$ and a strong cusp at $x_{\max}$. For the map of model II
the left derivative at the maximum is almost vertical. Again the appearance of the map
of model II is much smoother compared to the map of model I. 
Regarding the autocorrelation function we have the
opposite situation compared to the previous example. Within model I
(fig.3(b)) we obtain a relative error of $22 \%$ for the
autocorrelation data while model II (fig.3(d) provides
an error of $45 \%$. Obviously model I is advantageous in the present case (at least
for the present number of grid points chosen for the ansatz of $h_f$): it nicely
reproduces the oscillations of the correlation function.  

Finally we study the case that the autocorrelation function decays
algebraically with the exponent $\gamma=-2.5$, i.e. we encounter
the case of long-range correlations. The invariant density
is chosen to be uniform. Both resulting maps (fig.4(a,c)) possess a large
but finite derivative at $x=1$. The map of model I possesses an almost 
vertical derivative at a single point on its right branch. In contrast to
our second example (see above) this point does not coincide with the fixed point.
It is conjectured that this single point with almost vertical derivative
is responsible for the observed power law decay of the correlation function (fig4(b))
(see also ref.\cite{Hor}). Analogously the map of model II (fig.4(c))
shows a very well-pronounced vertical derivative at the cusp.
It is interesting to observe that in this case both
models I and II provide very satisfactory approximations to the
prescribed autocorrelation data (fig4(b,d)), i.e. 
a relative error of only $13 \%$. 

\section{Summary}

We have introduced a stochastic Monte-Carlo based approach to the inverse problem which uses
both the invariant density as well as a finite number of points of the
autocorrelation function as prescribed statistical quantities.
A key ingredient for this approach was a recently
found representation for one-dimensional complete chaotic and single-humped dynamical system
in terms of an auxiliary function $h_f$. This representation is a 
formal solution of the inverse Frobenius-Perron problem and provides the
dynamical system $f(x)$ explicitly as a function of the given measure and
the function $h_f$. $h_f$ therefore reflects the freedom of changing the map without 
changing its invariant measure (density). In order to quantify the freedom
available for the determination of $h_f$, i.e. to parametrize its functional
space, we have introduced two different models which allow to vary $h_f$ extensively.
The parameters involved are then used within our stochastic minimization
procedure to obtain a correlation function with least deviation from a 
prescribed autocorrelation signal. Through a number of examples we have demonstrated
that our approach possesses an enormous flexibility allowing for a large variety
of qualitatively different behaviour of the density and correlation function.
To our knowledge there is in general no unique map which belongs to a given density and correlation
function. This fact has to be seen in the context of the present investigation
as an advantage since it allows for a great flexibility and possible variety
with respect to the underlying dynamical systems.

We would like to mention that the above-discussed prescribed
behaviour of the autocorrelation data (exponential, oscillating, power law decay)
is, strictly speaking, enforced only for the first five points included in
our Monte-Carlo optimization. In principle it is imaginable that this
behaviour represents a transient and the asymptotics of the correlation
function might show a different behaviour. To determine exact asymptotic
properties of certain dynamical systems is however not the issue of the
present paper. Our goal is to extend the inverse problem by including the
correlation function in terms of a few (experimentally) available data points,
thereby enabling us to design a dynamical system with desired
statistical properties. Furthermore our approach might be suggestive in terms of influencing
or controlling dynamical systems \cite{Bollt,Gora} in a certain way. 

\section{Acknowledgements}
One of the authors (P.S.) thanks the Max-Planck Institute for Physics of Complex
Systems in Dresden for its kind hospitality. D.P. acknowledges financial
support by the Deutsche Forschungsgemeinschaft and the Landesgraduiertenf\"orderungsgesetz (LGFG).

{}
\vspace*{1.0cm}

\begin{center}
{{\bf{FIGURE CAPTIONS}}}
\end{center}

\noindent
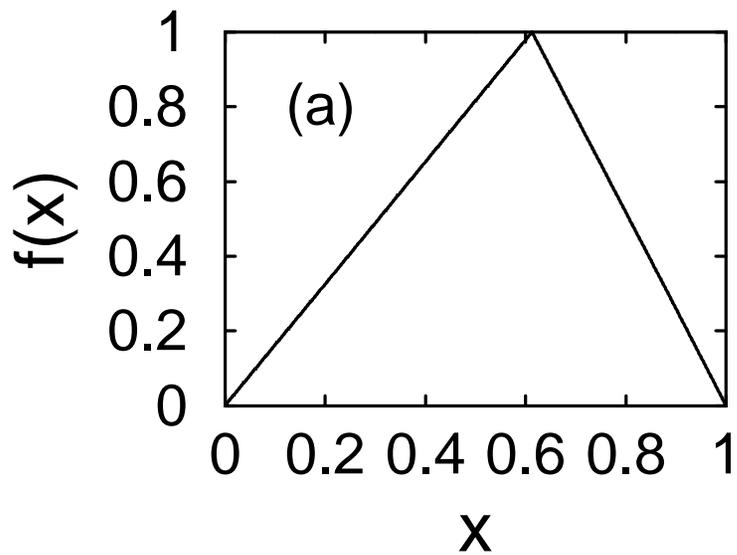
{\bf Figure 1:} The stochastic minimization results for a dynamical system with
uniform invariant density and exponentially decaying autocorrelation function.
(a) The resulting map using model I.
(b) The autocorrelation function for the map of model I (solid line) and
the corresponding prescribed data (full circles).
(c) The resulting map using model II.
(d) The autocorrelation function for the map of model II (solid line) and
the corresponding prescribed data (full circles). 
\vspace*{0.5cm}

\noindent
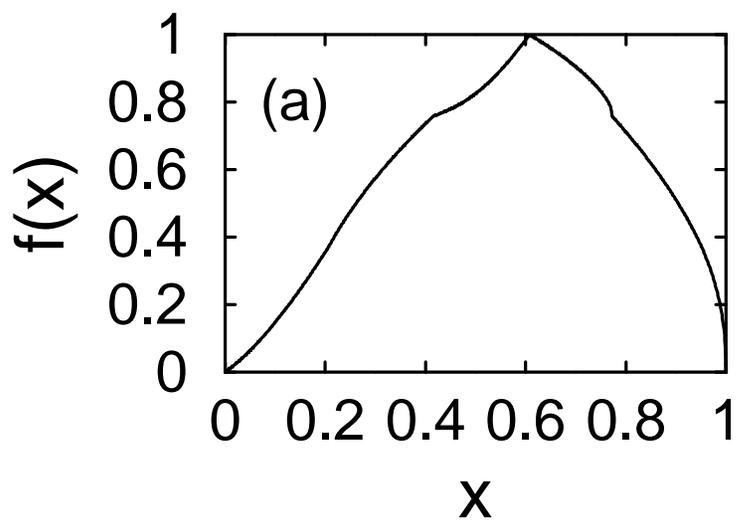
{\bf Figure 2:} Same as in figure 1 but for prescribed
linear invariant density and exponentially decaying autocorrelation function.
\vspace*{0.5cm}

\noindent
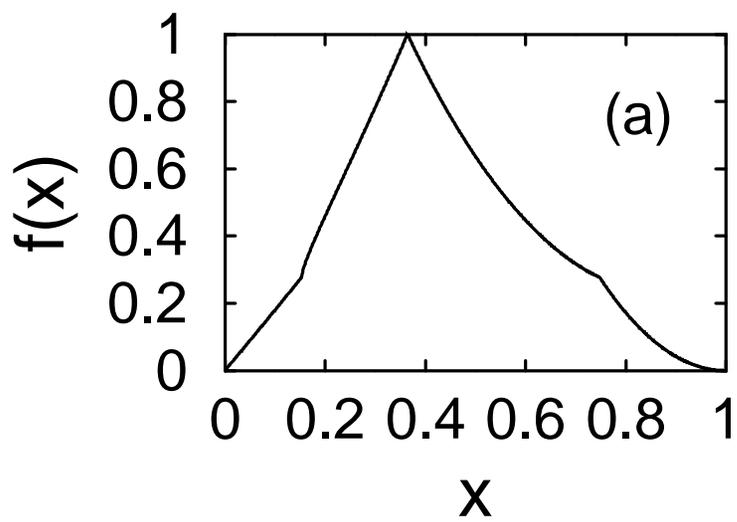
{\bf Figure 3:} Same as in figure 1 but for prescribed
power-law invariant density ($\beta=-0.5$) and oscillatory decaying
autocorrelation function.
\vspace*{0.5cm}

\noindent
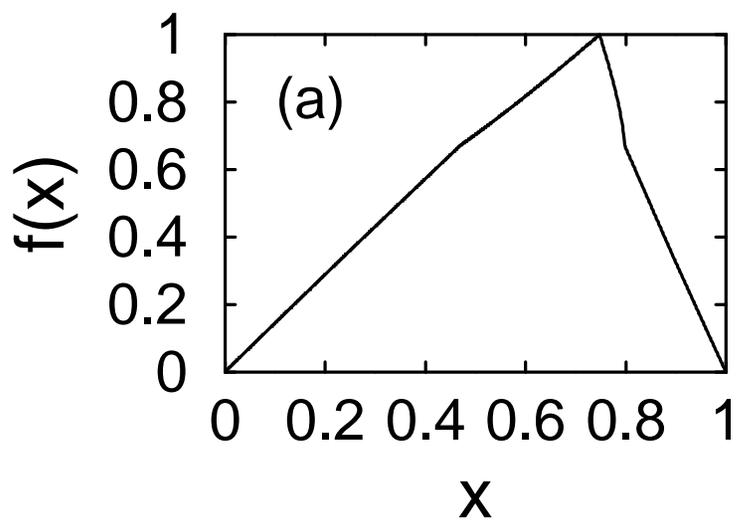
{\bf Figure 4:} Same as in figure 1 but for prescribed
uniform invariant density and power-law decaying autocorrelation function
(with exponent -2.5).

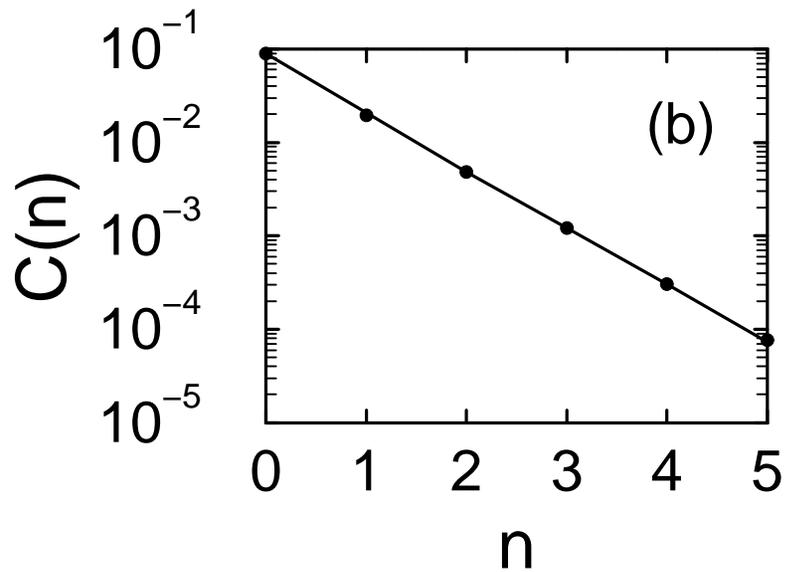
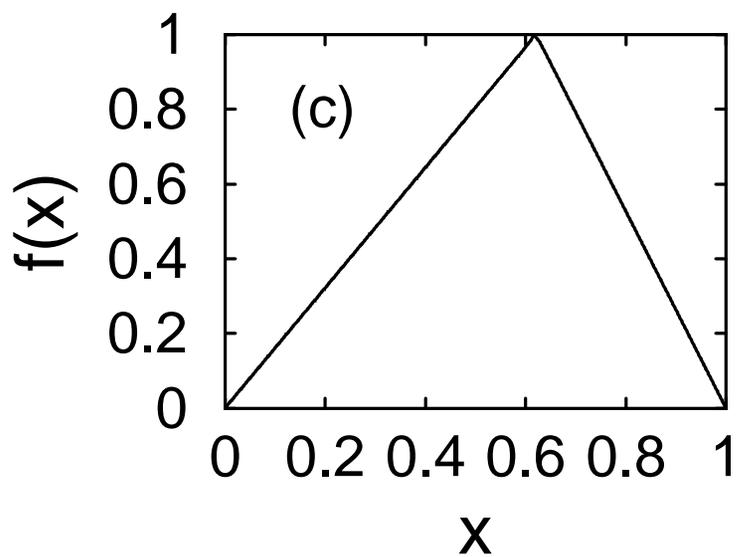
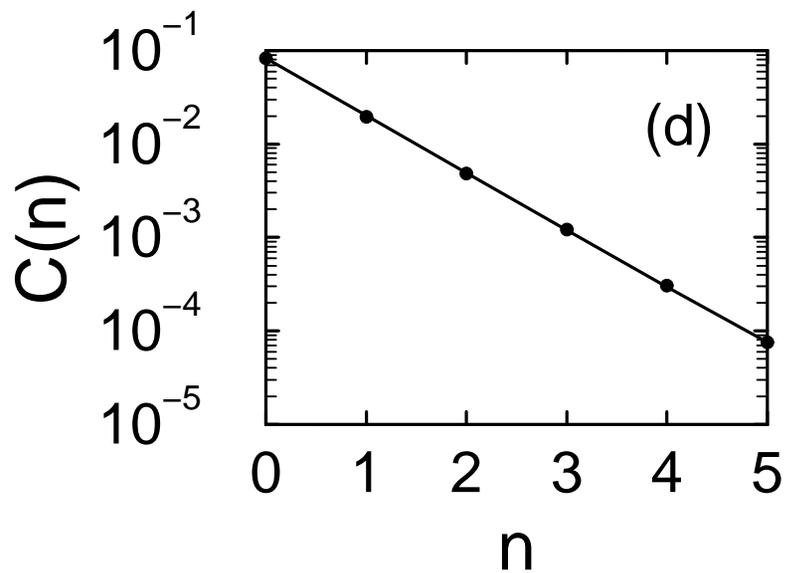
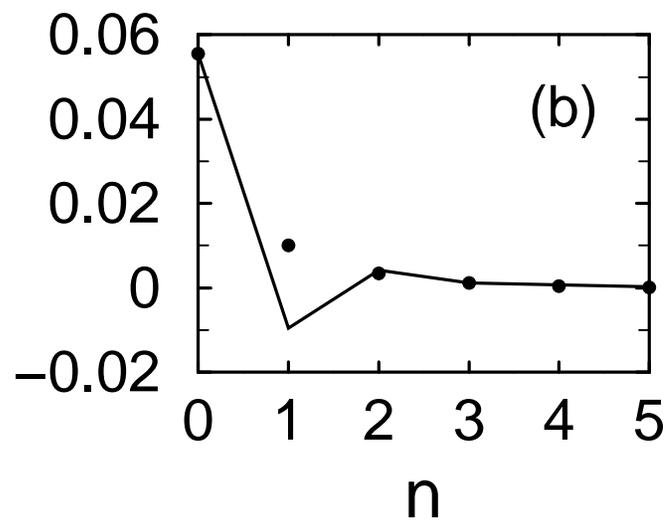
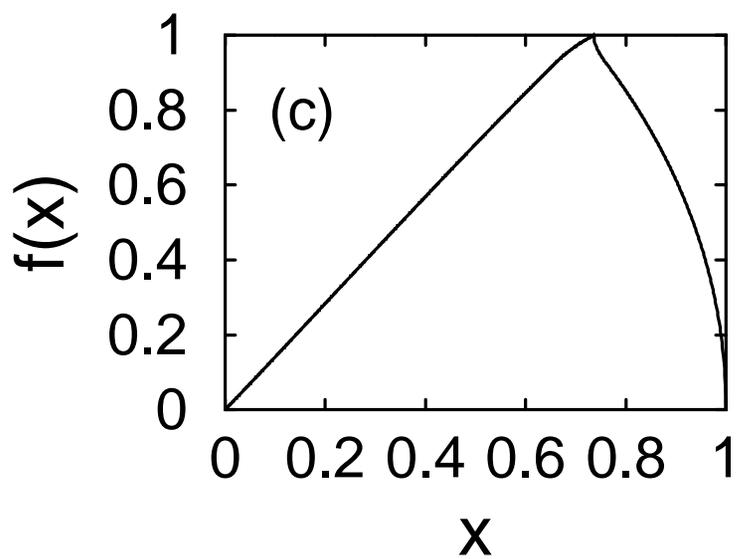
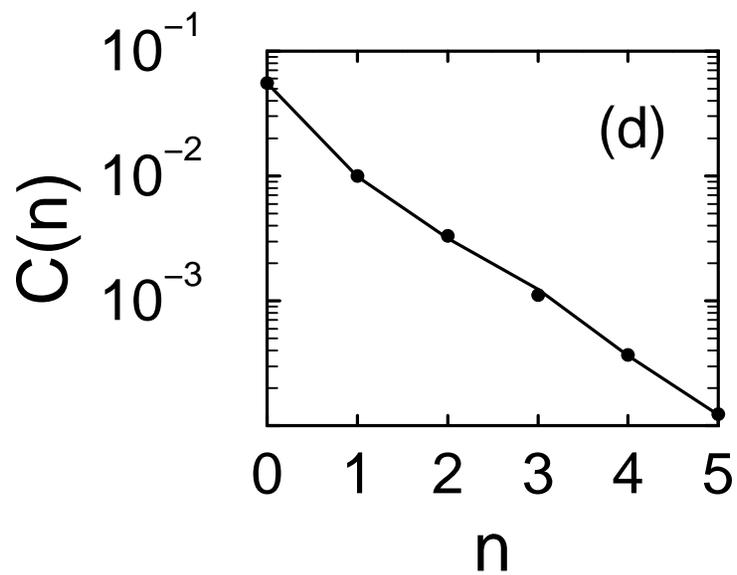
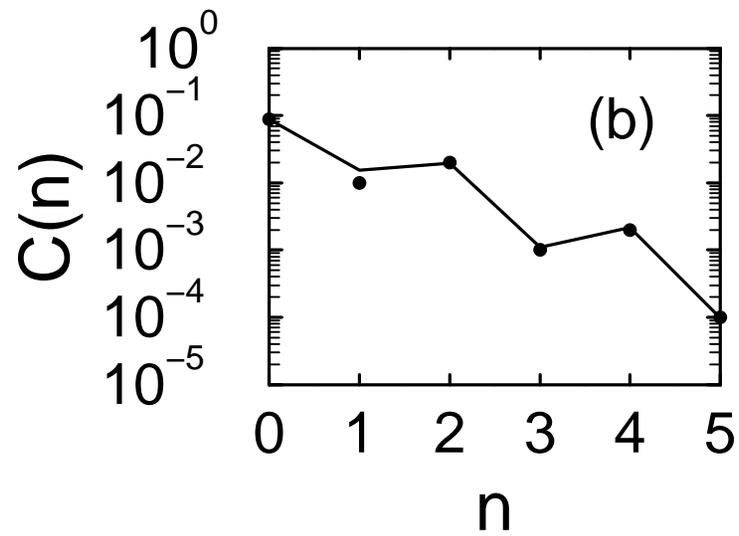
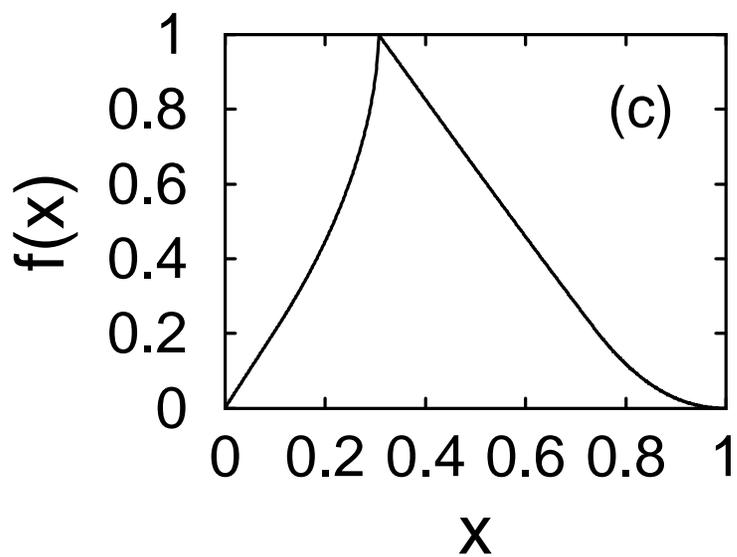
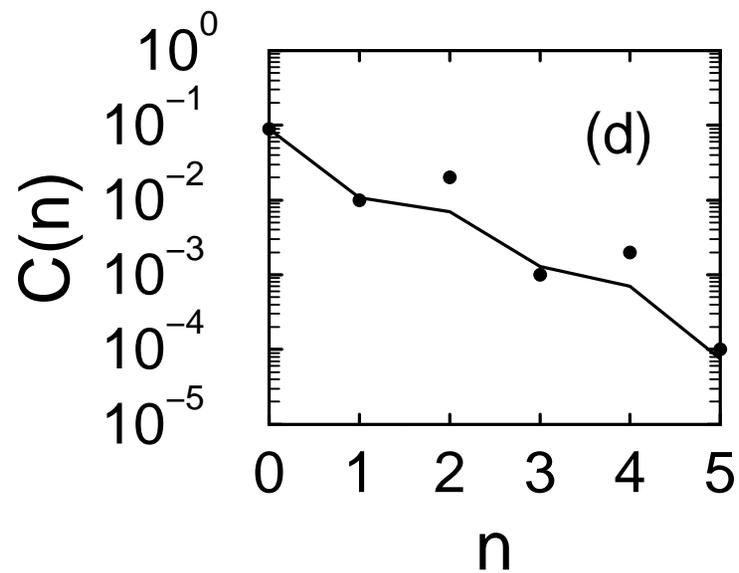
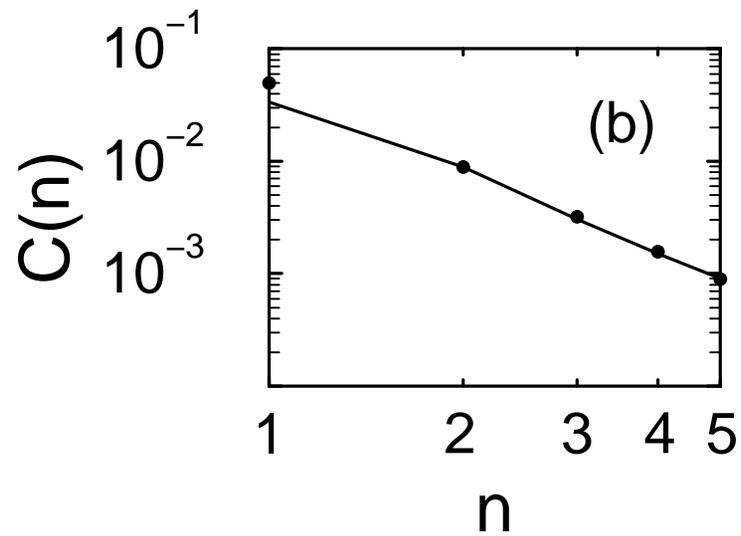
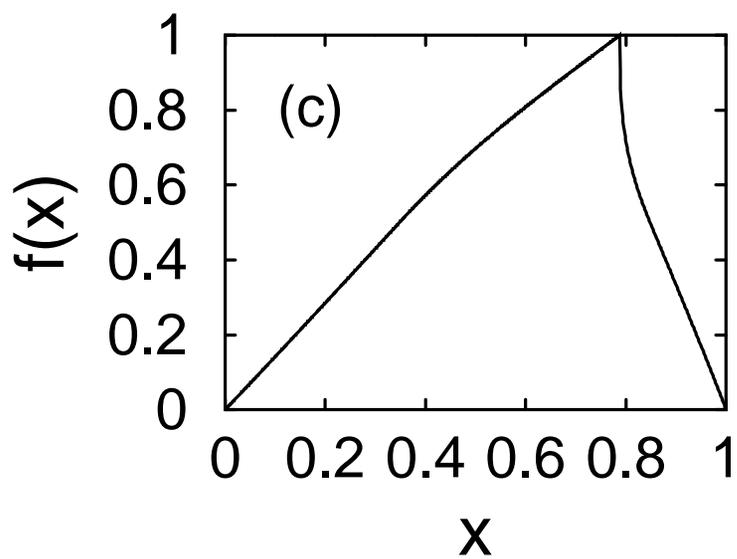
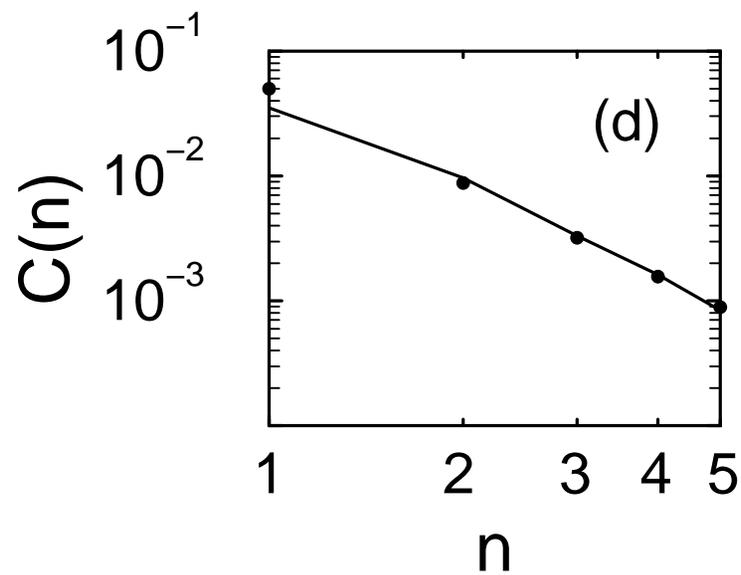
\end{document}